\documentclass[sigconf]{acmart}
\usepackage{makecell}
\AtBeginDocument{%
  }

\copyrightyear{2026}
\acmYear{2026}
\setcopyright{cc}
\setcctype{by-nc-nd}
\acmConference[CHI '26]{Proceedings of the 2026 CHI Conference on Human Factors in Computing Systems}{April 13--17, 2026}{Barcelona, Spain}
\acmBooktitle{Proceedings of the 2026 CHI Conference on Human Factors in Computing Systems (CHI '26), April 13--17, 2026, Barcelona, Spain}
\acmPrice{}
\acmDOI{10.1145/3772318.3791914}
\acmISBN{979-8-4007-2278-3/2026/04}

\begin{document}
\title[Rethinking User Empowerment in AI Recommender System]{Rethinking User Empowerment in AI Recommender System: Innovating Transparent and Controllable Interfaces}

\author{Mengke Wu}
\email{mengkew2@illinois.edu}
\affiliation{
    \institution{University of Illinois Urbana-Champaign}
    \department{School of Information Sciences}
    \city{Champaign}
    \state{Illinois}
    \country{USA}
}

\author{Weizi Liu}
\email{weizi.liu@tcu.edu}
\affiliation{
    \institution{Texas Christian University}
    \department{Bob Schieffer College of Communication}
    \city{Fort Worth}
    \state{Texas}
    \country{USA}
}

\author{Yanyun Wang}
\email{mia.wang@colorado.edu}
\affiliation{
    \institution{University of Colorado Boulder}
    \department{College of Communication, Media, Design and Information}
    \city{Boulder}
    \state{Colorado}
    \country{USA}
}

\author{Weiyu Ding}
\email{weiyu.ding@colorado.edu}
\affiliation{
    \institution{University of Colorado Boulder}
    \department{College of Communication, Media, Design and Information}
    \city{Boulder}
    \state{Colorado}
    \country{USA}
}

\author{Mike Yao}
\email{mzyao@illinois.edu}
\affiliation{
    \institution{University of Illinois Urbana-Champaign}
    \department{Institute of Communications Research}
    \city{Champaign}
    \state{Illinois}
    \country{USA}
}

\renewcommand{\shortauthors}{Wu et al.}

\begin{abstract}
AI-driven recommender systems are often perceived as personalization black boxes, limiting users’ ability to understand how their data shapes content (information asymmetry) or to influence system behavior meaningfully (power asymmetry). This study explores how design can strengthen user agency by integrating transparency with actionable control. We developed a \textit{provotype} that introduces new interface features for managing data use, discovering varied content, and configuring context-based recommending modes. The walkthroughs and interviews with 19 participants show how these features help users interpret personalization signals, understand how their actions influence outcomes, address concerns from unwanted inference to narrow feeds (e.g., filter bubbles), and build trust in the system. We also identify strategies for promoting adoption and awareness of agency-enhancing features. Overall, our findings reaffirm users’ desire for active influence over personalization and contribute concrete interface mechanisms with empirical insights for designing recommender systems that foreground user autonomy and fairness in AI-driven content delivery.
\end{abstract}





\maketitle

\section{INTRODUCTION}
Behavioral targeting and predictive algorithms have revolutionized information dissemination by enhancing efficiency and transforming how media is produced and distributed. These systems draw on user data to anticipate their interests and automatically curate relevant content \cite{pandey2011learning, yoneda2019algorithms}, and now operate across social media, news platforms, forums, streaming services, e-commerce, and more. While much research has examined the development and impact of these recommender systems (RS), concerns remain about their effects on human-AI interactions and the pursuit of human-centered design \cite{ehsan2021operationalizing, shin2019role}. One central critique is the opacity of algorithmic decision-making, often portrayed as a “black box” where users can only access and consume finalized outputs \cite{pasquale2015black, burrell2016machine}, lacking explanations and involvement \cite{felzmann2020towards, weller2019transparency}. This produces two forms of imbalance: \textbf{1) information asymmetry}, which limits users’ perceptual agency by concealing the mechanisms behind recommendations \cite{kizilcec2016much, rader2018explanations, sonboli2021fairness}, and \textbf{2) power asymmetry}, which erodes users’ behavioral agency by restricting their ability to shape or intervene in algorithmic decisions \cite{cheng2023overcoming, molina2022ai, sundar2020rise}. These asymmetries challenge user-centered design and risk manipulation, information blockage, and loss of trust \cite{conover2011political, kizilcec2016much, nguyen2014exploring, ruhr2023intelligent}.

Advances in AI transparency, explainable AI (XAI) \cite{kizilcec2016much, rader2018explanations}, and human-AI collaborative decision-making (HACD) have aimed to mitigate these problems. For example, cookie disclaimers reduce information asymmetry by making data practices more visible and manageable \cite{segijn2021literature}, while permission controls \cite{burrell2016machine, segijn2021literature} and collaborative input mechanisms \cite{lai2023towards, molina2022ai} shift some power to users. However, these approaches often rely on partial or rigid controls (e.g., simple on/off), limiting finer adjustments or more diverse controls, such as how different data sources should contribute to recommendation logic or how strongly personalization should be applied.

As one concrete consequence of these asymmetries, RS often produces what are described as \textit{“filter bubbles,”} where personalization amplifies already-observed behaviors and narrows exposure to alternatives \cite{pariser2011filter, aridor2020deconstructing}. Prior work has attempted to counter these effects by introducing diversity, novelty, or serendipity \cite{kaminskas2016diversity, castells2021novelty, ziarani2021serendipity}, but these solutions remain backend-centric. Recent work also highlights a broader tension between users’ stated preferences and system-inferred preferences, where people articulate broader desires (e.g., diversity or longer-term value) that are not captured by engagement signals \cite{rezk2024agency,loepp2014choice}. This misalignment also shows that recommendation dynamics, including those that produce filter bubbles, arise under limited agency: users cannot understand or meaningfully adjust the forces that influence their feeds. Together, these issues underscore the need for RS designs that allow users to specify the kinds and degrees of tailored recommendations they prefer from algorithmic predictions.

Building on these gaps, this paper pursues two goals: \textbf{1) to explore how interface-level features can enhance transparency and control in recommendation processes}, and \textbf{2) to study how such design interventions affect user perceptions, attitudes, and sense of agency (the ability to understand, evaluate, and meaningfully influence) toward the system.} We designed a \textit{provotype} (provocative prototype) that lets users adjust data practices and navigate content discovery. Participants interacted with the provotype and shared feedback in semi-structured interviews. We use this process as a generative tool to identify design considerations that, in turn, inform broader challenges. The findings contribute to ongoing debates on algorithmic fairness, user autonomy, and human–AI collaboration, while offering practical mechanisms for RS that balance algorithmic efficiency with user agency. Ultimately, this work advances a vision of RS that fosters understanding, control, and trust in the digital ecosystem.

\section{RELATED WORK} \label{lit}

\subsection{Information Asymmetry} 
Information asymmetry in algorithm-based RS refers to the discrepancy between what users can observe (the recommendations) and what remains hidden (how recommendations are generated) \cite{lepri2021ethical, eslami2019user, sonboli2021fairness}. This asymmetry often prevents users from fully comprehending system logic and undermines perceptual user agency, standing in contrast to the pursuit of “interpretability” and “explainability” in complex algorithms. Prior work stresses the importance of making RS transparent and interpretable from the user's perspective \cite{cramer2008effects, kizilcec2016much, sonboli2021fairness}, and connects this to situation awareness in autonomous and AI systems for effective interaction and oversight \cite{endsley2023supporting}. A range of strategies has been proposed to address this issue: 1) source transparency, clarifying why and where such recommendations originate \cite{felzmann2020towards}; and 2) process transparency, often realized through Explainable AI (XAI) to elucidate system behavior \cite{rader2018explanations, ras2018explanation}. Common mechanisms, such as Cookie disclaimers, serve as forms of transparency by informing users about what data is collected and how it contributes to personalization \cite{millett2001cookies, segijn2021literature}.

Guidelines have also been proposed to make these systems more transparent and usable. Liao et al. introduced a “question bank” to uncover user needs around explanations \cite{liao2020questioning}, while Microsoft's HAX framework offers structured guidance for building user-centered XAI \cite{amershi2019guidelines}. Broader usability work in user interface and XAI design further advocates to address needs for learning “what to explain” and “how to explain” \cite{eiband2018bringing, wolf2019explainability}.

\subsection{Power Asymmetry}
While information asymmetry primarily limits perception, power asymmetry constrains agency behaviorally. In AI-driven RS, power asymmetry reflects an imbalance between user control and algorithmic authority \cite{stefanija2023power, konig2024challenges}. Prior work describes an inherent tension between human and machine agency, noting that the loss of agency drives fears about automation and underscoring the importance of human-AI synergy, in which users directly shape algorithms to match their needs \cite{sundar2020rise}. In contrast, persistent power asymmetry can reduce system adoption, as lower perceived control diminishes both perceived usefulness and engagement intentions \cite{chen2018app}. Proactive and reactive relationships between users and algorithms have also been discussed: proactive interactions involve machine agency by predicting user needs and delivering content based on analyzed patterns, whereas reactive interactions exhibit user agency through explicit user requests that drive personalized suggestions \cite{zhang2019proactive}.

\subsubsection{Existing User Control in RS}
User control mechanisms are essential to mitigate power asymmetry and improve transparency, trust, satisfaction, and ethical concerns such as accountability and democratic participation \cite{harambam2019designing}. Emerging ideas like \textit{interactive transparency} \cite{molina2022ai} and human-AI collaborative decision-making (HACD) \cite{lai2023towards, schemmer2022meta} stress respecting user feelings and involving them in algorithmic decision-making to bolster trust and perceived control.

\begin{figure*}
    \centering
    \includegraphics[width=1\linewidth]{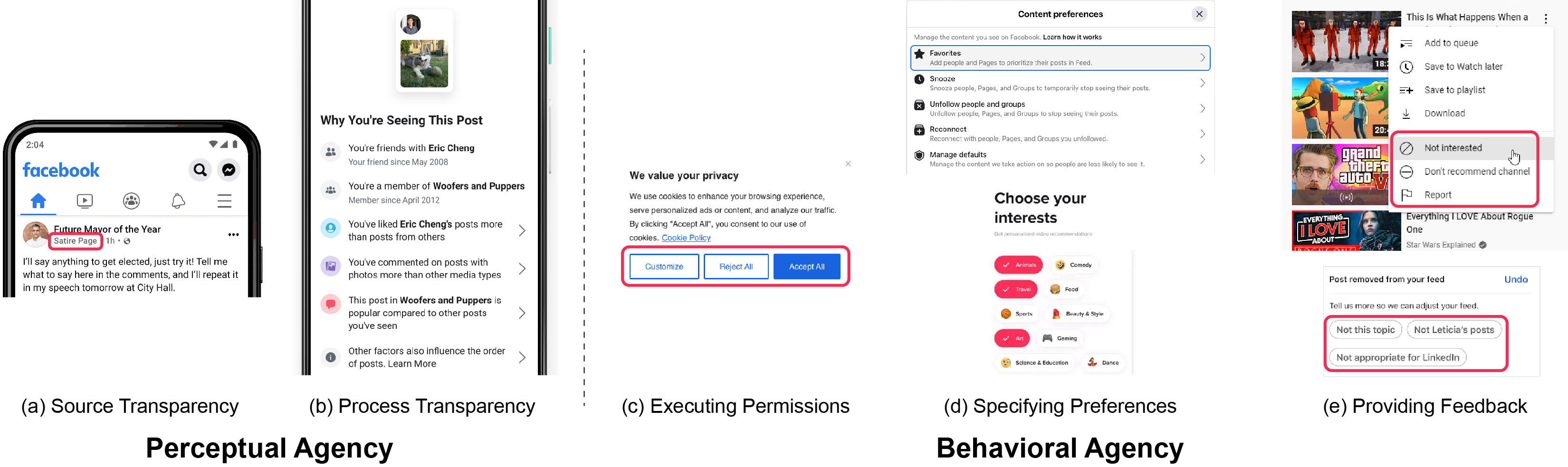}
    \caption{Examples of Different Types of Agency. Perceptual Agency (Left): (a) Source Transparency, (b) Process Transparency; Behavioral Agency (Right): (c) Executing Permissions, (d) Specifying Preferences, (e) Providing Feedback.}
    \label{fig:current}
\end{figure*}

Behavioral agency commonly manifests in three ways: 1) granting or withholding permissions (e.g., declining cookies) \cite{berens2024cookie, segijn2021literature}, 2) specifying preferences for personalized recommendations \cite{rashid2002getting, zheng2022perd}, and 3) providing feedback to refine algorithmic decisions \cite{cheng2023overcoming, molina2022ai}. For example, studies have enabled users to understand and adjust how their data influences recommendations by inspecting and modifying their profiles, thereby boosting transparency and trust \cite{bakalov2013approach}. This has expanded into interfaces that support preference specification, recommendation adjustments, and feedback, creating more interactive and responsive experiences \cite{jannach2017user, jannach2019explanations}. Other efforts include adjustable parameters (e.g., popularity, recency), immediate control over outputs, and enhanced satisfaction \cite{harper2015putting}. Meta-recommender systems combine multiple recommendation sources to further empower users to personalize and control the recommendation process \cite{schafer2002meta}. Figure \ref{fig:current} presents a range of user agency in current practices.

Research also examines how users perceive control and the psychological factors behind it. Value-added functions, such as tailoring control interfaces to individual traits, can increase recommendation acceptance \cite{jin2020effects}, while pairing control with visual explanations further enhances transparency and satisfaction \cite{tsai2021effects}. In conversational RS, adaptability, understanding, and responsiveness are critical for fostering a sense of control \cite{jin2021key}. Overall, well-designed mechanisms can reduce cognitive load and improve engagement, directly strengthening users’ perceived control \cite{laban2020effect}.

\subsection{Filter Bubbles \& Preference Misalignment} \label{filter}
As RS become increasingly optimized for predictive accuracy (e.g., \cite{wu2022survey}), concerns grow that over-personalization produces \textit{“filter bubbles”}: algorithms infer preferences from past behavior, rank content accordingly, and elicit further engagement that reinforces those same patterns. This feedback loop can narrow exposure, suppress alternatives, and heighten vulnerability to misinformation by removing countervailing sources \cite{pariser2011filter, nguyen2014exploring, adamopoulos2014over}. For users, this often leads to predictable or stagnant feeds and fewer opportunities for exploration \cite{aridor2020deconstructing}. HCI and RS communities have long cautioned that accuracy alone can mask experiential isolation while inflating satisfaction metrics \cite{zhang2024see, wang2022user}.

\subsubsection{Mitigation, Limitation, and Implication}
To address beyond-accuracy objectives, earlier work introduced algorithmic diversity, novelty, and serendipity \cite{ziarani2021serendipity, kaminskas2016diversity, castells2021novelty}. \textit{Novelty} captures how much a recommendation departs from a user's prior consumption, while \textit{diversity} concerns intra-list heterogeneity across categories, styles, or attributes at a given moment, independent of user familiarity. Yet, neglecting surprise can still lead to monotonous outcomes, and the assumption that users always prefer familiar content can be flawed \cite{iaquinta2008introducing}. Therefore, \textit{serendipity} adds unexpectedness with relevance to introduce novelty with minimal accuracy trade-offs \cite{ziarani2021serendipity}.

However, most efforts remain backend-centric: they tune system outputs without giving users control over how much or what kind of variation they want. This limitation contributes to a broader structural tension between users’ stated preferences and the revealed preferences inferred from behavioral traces (e.g., clicks, dwell time, scrolling). Studies in news personalization, interactive preference elicitation, and intimate platforms show that users often express long-term, value-driven goals for diversity, fairness, and cross-cutting viewpoints. Still, recommender logic tends to prioritize past engagement signals and override these aspirations \cite{rezk2024agency,loepp2014choice, hutson2018debiasing}. As a result, even users who seek broader or exploratory content have limited means of directing it through the interface.

This shows how one-way recommendation processes, including those that produce filter bubbles, exemplify a concrete consequence of limited agency: systems infer preferences and push content without users’ explicit direction, leading to constrained exposure and outputs that feel misaligned with users' values or interests \cite{wang2022user}. One attempt to bridge this gap is conversational RS, which lets users articulate intentions or preferences via natural dialogue, reducing reliance on behavioral traces and granting more explicit control \cite{jannach2021survey, cai2020predicting}. However, conversational RS often requires sustained conversational effort and is better suited to task-oriented item search than casual browsing \cite{gao2021advances}. Our work complements this line by exploring lightweight, interface-level mechanisms for steering how recommendations deviate from AI-driven curation in everyday use, without the interaction load or proactive querying.

\subsubsection{Information Fairness}
A related concern is user-side information fairness \cite{deldjoo2021flexible}. Collaborative-filtering pipelines often overlook unobserved attributes and emerging user groups, meaning that some individuals are systematically exposed to fewer catalog choices even when they should have access to comparable options \cite{misztal2021bias}. As user data accumulates and models grow more complex, users will inevitably encounter divergent slices of content and may never see the “whole picture.” To address this, mixing a small proportion of non-personalized or even provocative items into otherwise tailored slates could broaden exposure, enhance information fairness, and strengthen choice confidence by surfacing items not yet signaled by the user. This strategy parallels work on exposure fairness via randomized ranking to equalize opportunity \cite{singh2018fairness, ubaid2021impact}, but shifts the locus of control to the interface, enabling users, not only backend models, to invoke broader perspectives when desired. To date, however, such front-end mechanisms remain uncommon, highlighting design opportunities and motivating our research.

\section{PROVOTYPE DESIGN}

\subsection{What is a \textit{Provotype} and Why We Use It}
How can we enhance user empowerment in AI RS? To explore this, we created features on both settings and feed pages for more explicit information disclosure and user intervention than typical RS, and embedded them into a \textit{provotype}: a provocative design artifact meant to spark discussion rather than move toward final production \cite{to2022interactive}. Unlike standard prototypes, which imply refinement and usability testing, a \textit{provotype} “feels out the boundaries of a problem” and encourages participants to question assumptions. This draws on IDEO’s notion of \textit{“sacrificial concepts”} that are early-stage ideas used to provoke conversations and help imagine alternative futures \cite{carey2009tangible}. Prior work shows the value of \textit{provotypes} in surfacing tensions and supporting co-creation by providing concrete ideas to respond to rather than abstract questions \cite{brandt2012tools, boer2012provotypes}. As such, our \textit{provotype} serves two goals: 1) gather user feedback on proposed features, and 2) act as a tangible probe to prompt reflection and discussion about user agency and control in RS.

We chose TikTok as the display platform for the \textit{provotype} due to its global popularity and diverse content ecosystem (e.g., lifestyle, learning, news, entertainment) \cite{boeker2022empirical, alley2022long}, which can facilitate engagement and allow participants to consider a wider range of content scenarios. Nevertheless, as core AI recommendation mechanisms are similar across platforms, TikTok served solely as a visual scaffold, not to imply the features are limited to it or social media. Wherever applicable, we translated descriptions and findings into insights that generalize beyond social media.

\subsection{Ideation Process}
\begin{figure*}
    \centering
    \includegraphics[width=1\linewidth]{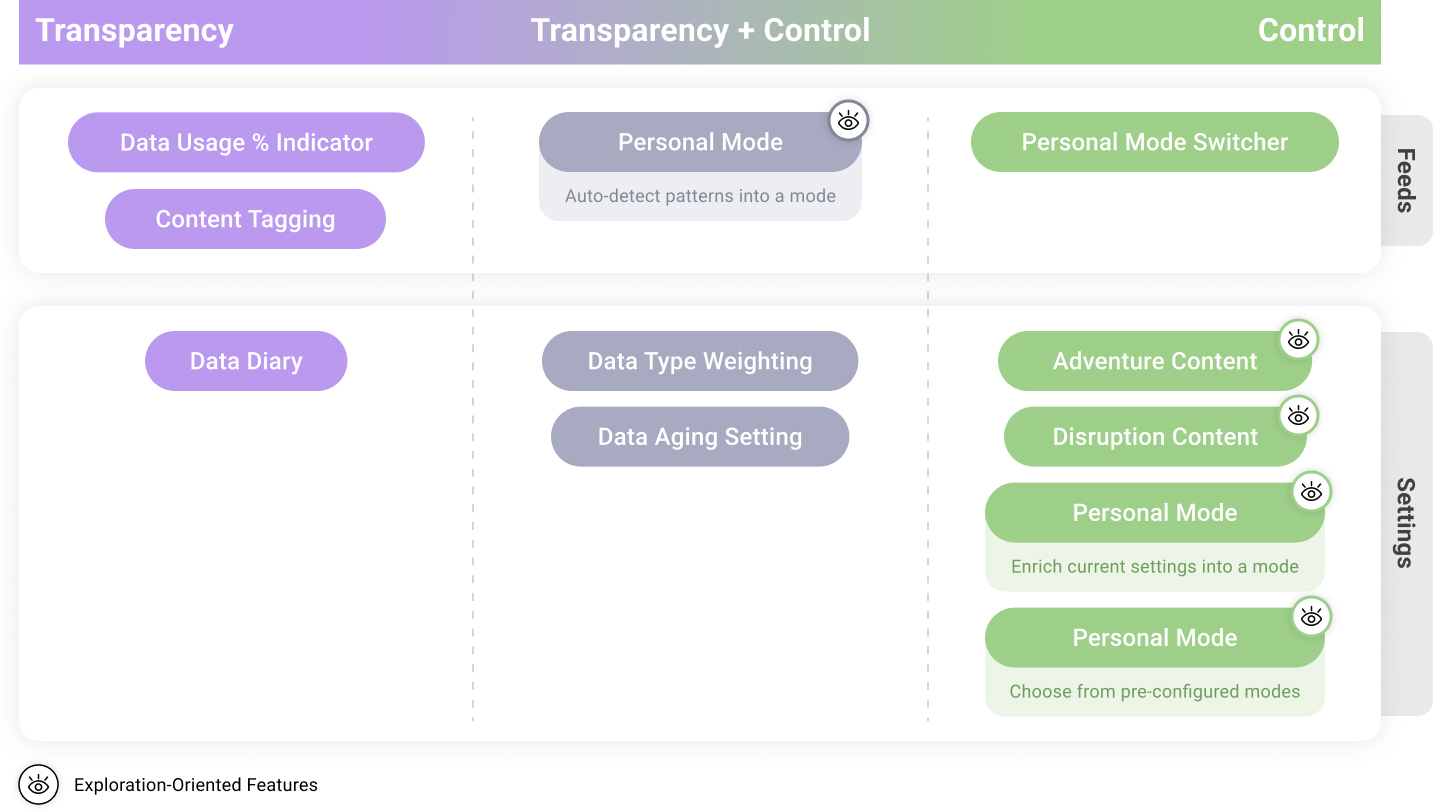}
    \caption{Feature Mapping by Focus and Placement.}
    \label{fig:map}
\end{figure*}

Our ideation process followed a structured, literature-informed approach. The team consisted of three researchers with expertise in human–AI interaction and recommender systems, including one with a solid UX design background. We grounded our exploration in the conceptual links outlined earlier: transparency features address information asymmetry, control features address power asymmetry, and exploration-oriented features target exposure narrowing as a downstream effect. Guided by this framing, we organized our ideating canvases around two central aspects of RS: data collection and personalization. Within each canvas, we reviewed literature-identified problems, drew inspiration from prior design research (e.g., \cite{storms2022transparency, luria2023co, wu2025negotiating, bemmann2024privacy}), and examined existing RS affordances to propose new countermeasures. We also considered where each feature should live, either directly on the feed page for in-situ interaction or in settings for more deliberate customization. All proposed features were then collectively evaluated within the research team against criteria like expected user impact, novelty, understandability, and usability. To reduce participant cognitive load in a one-time study, the highest-scoring features from each canvas were selected for the provotype. Figure \ref{fig:map} shows how these implemented features map across focus (transparency vs. control, with exploration-oriented features labeled) and placement (feed-level vs. settings-level), with Section \ref{design} detailing each.

\subsection{Proposed Features in the \textit{Provotype}} \label{design}
We implemented the provotype in Figma with interactivity to simulate a real application. It supported clicking all essential chips and performing other gestures (e.g., drag sliders, hover to view information, scroll through entries), which effectively conveyed the visual structure and user flow of all features.

\subsubsection{Data Management}
\begin{figure*}
    \centering
    \includegraphics[width=1\linewidth]{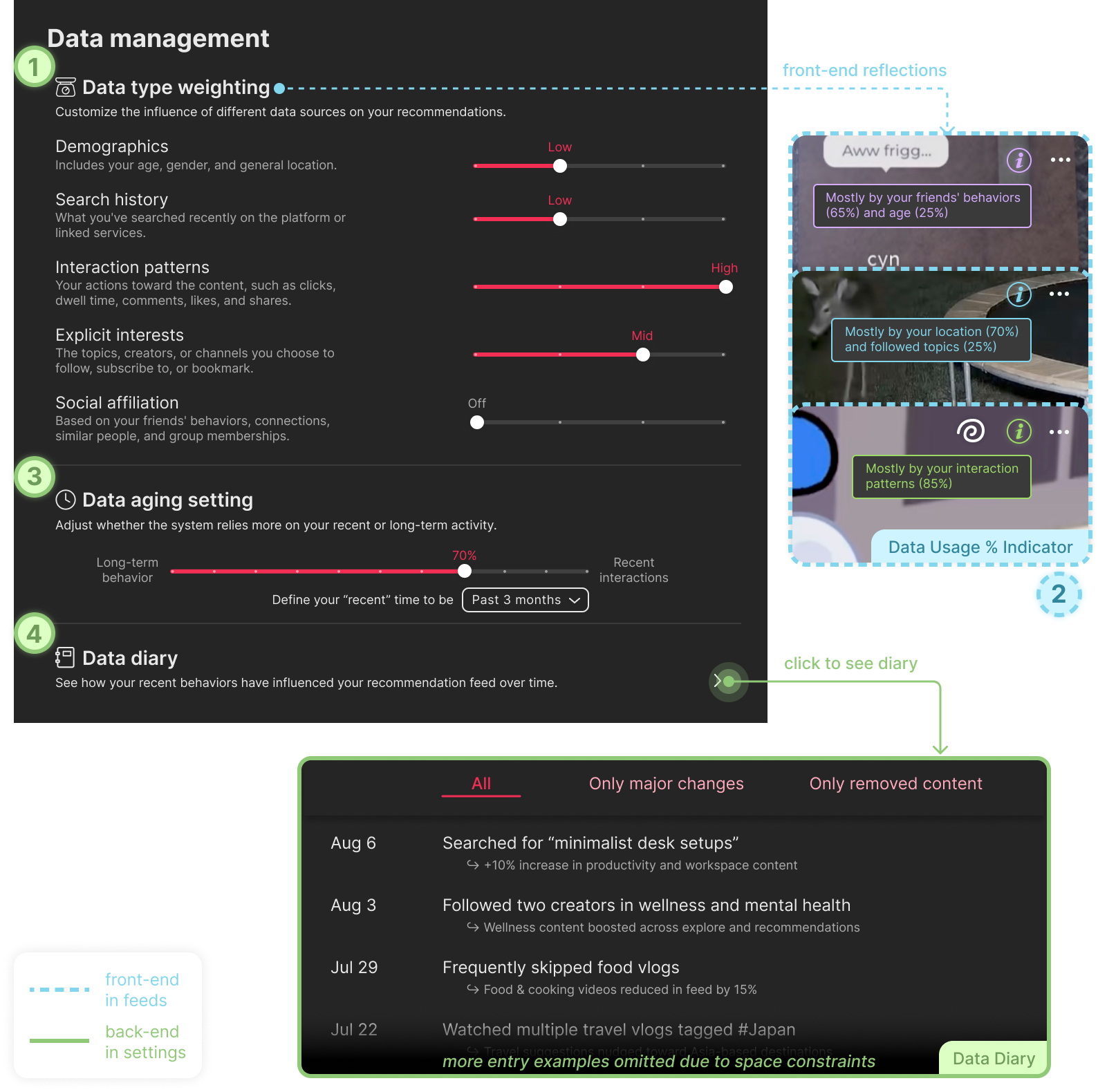}
    \caption{Data Management Suite: 1) Data Type Weighting, 2) Data Usage \% Indicator, 3) Data Aging Setting, 4) Data Diary.}
    \label{fig:dm}
\end{figure*}

The Data Management suite addresses information and power asymmetries by making hidden algorithmic inputs visible and adjustable. Prior work shows that users gain understanding and confidence when they can see how their data influences recommendations \cite{rader2018explanations, sonboli2021fairness}, yet existing platforms typically offer only coarse controls and high-level explanations. Drawing from prior work \cite{wang2024chaitok, bemmann2024privacy, storms2022transparency, luria2023co}, this suite extends actionable transparency by enabling more holistic and fine-grained manipulations of data types, temporal scopes, and observable system responses. Figure \ref{fig:dm} illustrates the suite and its functions.

\textbf{Data Type Weighting:} Users can adjust the influence of five major data categories on their recommendations: demographics (age, gender, location), search history (recent queries on the platform or linked services), interaction patterns (e.g., clicks, dwell time, comments, likes, shares), explicit interests (followed topics, creators, or channels), and social affiliation (friends’ activities and similar user profiles). While existing platforms offer some control on clearing watch history (e.g., YouTube) or toggling broad personalization settings (e.g., TikTok), they often operate as blunt on/off switches. Inspired by prior slider-based designs \cite{wang2024chaitok, bemmann2024privacy, storms2022transparency}, we extend the affordances by enabling finer control over valuing or devaluing different data types (see Figure \ref{fig:dm}-1). This targeted control helps align user intentions with algorithmic signals, reducing reliance on irrelevant or sensitive data.
    
\textbf{Data Usage \% Indicator:} Building on the Data Type Weighting, this component visualizes how they shape individual recommendations. Existing \textit{“Why you see this”} explanations (e.g., Facebook, TikTok) offer only high-level categories and do not convey the strength of contributing signals or show how user adjustments alter system behavior. Drawing on the “Plug and Play” in \cite{luria2023co}, we attach a color-coded “info” icon to each item to surface the dominant data source(s) and their approximate contribution when users hover over it (e.g., blue = demographics, green = interaction patterns) (see Figure \ref{fig:dm}-2). To reduce cognitive load, we omit full numeric breakdowns of all contributing types. This creates a direct feedback loop between backend weighting sliders and front-end feed: for example, users should see more green icons after raising Interaction Patterns to “high,” making user impact visible and reinforcing their sense of agency over how the system operates.
     
\textbf{Data Aging Setting:} Beyond data types, users can adjust the time-wise scope of data used for personalization. While current systems allow users to edit or delete history, or set how long the history is retained (e.g., YouTube, Facebook), the Data Aging Setting introduces a new temporal dimension by letting users specify whether recommendations should rely more on recent or long-term behavior (e.g., prioritize activities from the past few days) (see Figure \ref{fig:dm}-3). By making the system’s handling of temporal data explicit and adjustable, this feature enhances transparency and helps users tailor recommendations to their evolving interests and needs.
    
\textbf{Data Diary:} Most platforms show activity histories but do not link user actions to changes in recommendation patterns. The Data Diary fills this gap by presenting a lightweight log that reveals how user actions (e.g., searches, followed creators, content skips) alter the type and frequency of recommended content over time (see Figure \ref{fig:dm}-4). This causal visibility echoes transparency efforts but extends them by illustrating how user behavior amplifies or dampens recommendations. For privacy and relevance, the diary retains only recent activity (e.g., three months), offering a concise and interpretable view without overwhelming historical details.

\subsubsection{Content Discovery}
\begin{figure*}
    \centering
    \includegraphics[width=1\linewidth]{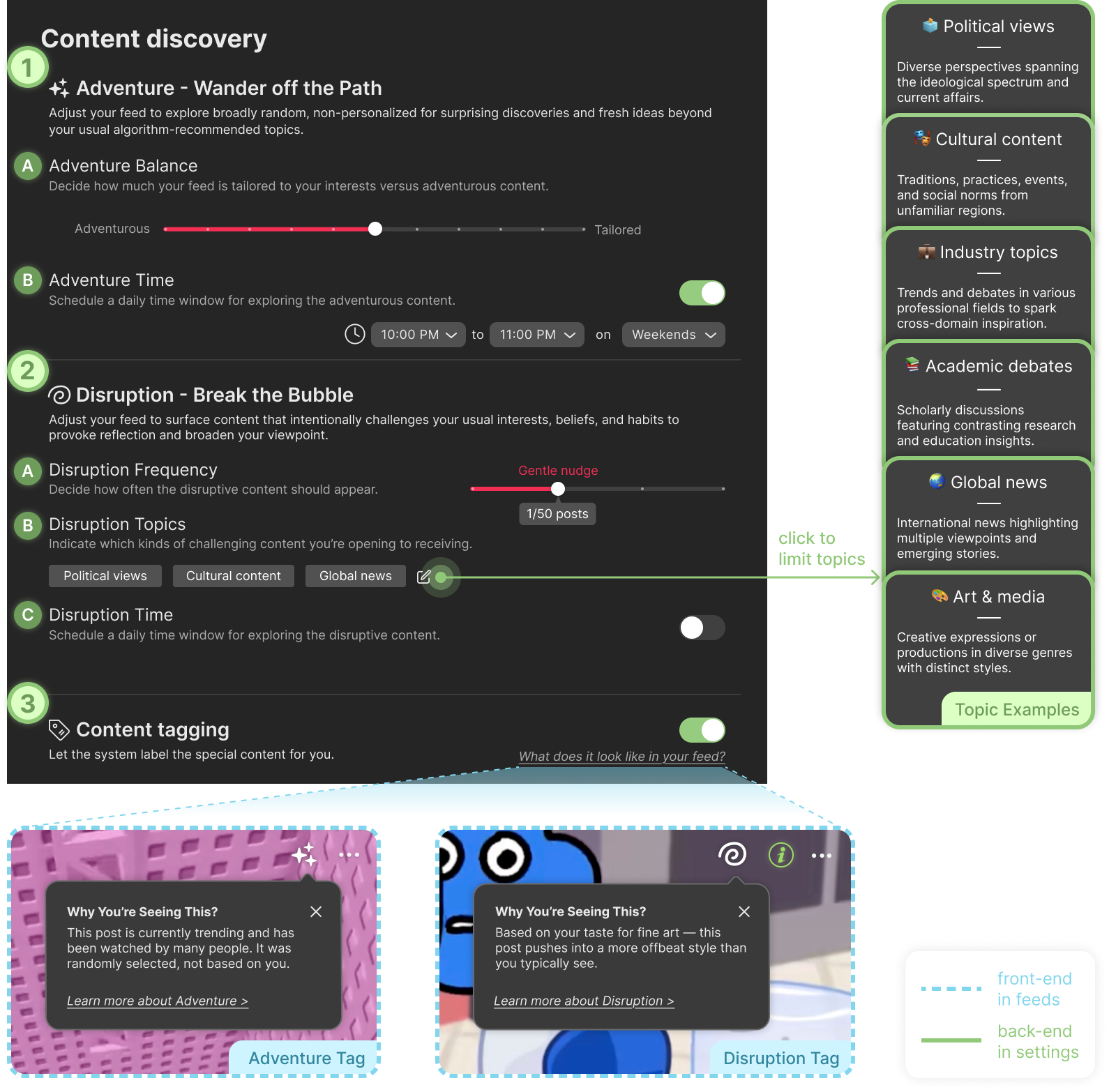}
    \caption{Content Discovery Suite: 1) Adventure Content (A - Balance, B - Time), 2) Disruption Content (A - Frequency, B - Topic, C - Time), 3) Content Tagging.}
    \label{fig:cd}
\end{figure*}

The Content Discovery suite supports intentional exploration to counter filter bubbles. Prior research highlights the value of diversified content for mitigating narrow exposure \cite{kaminskas2016diversity, singh2018fairness}, but these efforts typically occur on the algorithmic side and limit users' visibility or control. This suite introduces a user-driven exploration framework that helps users encounter new perspectives, broaden their interests, and reflect on habitual consumption patterns, allowing them to consciously guide what, when, and how exploratory content appears in their feed. Figure \ref{fig:cd} illustrates the suite and its functions.

\textbf{Adventure Content:} This function invites users to “wander off the path” from their tailored feeds and encounter random, non-personalized materials for fresh ideas and unexpected discoveries. Many platforms provide exploration spaces, such as TikTok's “Explore”, YouTube's “Trending”, or Spotify's “Discovery,” but these typically require leaving the main feed. Even when systems diversify recommendations \cite{ziarani2021serendipity, kaminskas2016diversity, castells2021novelty}, such interventions operate invisibly within algorithms and do not allow users to control how such content appears. To transfer this power to users, Adventure introduces two mechanisms: 1) The Adventure Balance slider adjusts the proportion of personalized versus random content from a fully tailored feed (right end) to a non-data-driven experience similar to visitor browsing or incognito mode (left end). Intermediate positions expose users to a mix of new perspectives while maintaining some familiarity (see Figure \ref{fig:cd}-1A). 2) The Adventure Time option adds further flexibility by letting users schedule designated windows for exploration, such as “show 30\% adventurous content on weekends from 10–11 p.m.” (see Figure \ref{fig:cd}-1B). Together, these controls provide a structured and proactive way to break out of content bubbles while preserving the benefits of personalization.
    
\textbf{Disruption Content:} Whereas Adventure introduces randomness, Disruption takes a more intentional approach to “break the bubble” by delivering content that challenges users’ typical interests, beliefs, or habits. Unlike random exploration, Disruption still uses user data but selects diverging content to provoke reflection and broaden perspectives. The strong name \textit{Disruption} signals that it may surface uncomfortable or even offensive content (e.g., political debates or opposing views), prompting more deliberate engagement. Users can adjust the Disruption Frequency from completely off (for those who prefer no disturbance) to gentle nudges (one in fifty posts), noticeable mixes (one in ten), or bold shifts (every post diverging from the usual feed) (see Figure \ref{fig:cd}-2A). To maintain comfort and agency, disruption can be limited to specific topics, such as political views, cultural content (e.g., from unfamiliar regions), or industry topics (for cross-domain inspiration), allowing users to choose where they welcome challenges (see Figure \ref{fig:cd}-2B). Similar to Adventure, users can schedule Disruption during particular windows (see Figure \ref{fig:cd}-2C). These mechanisms provide meaningful, user-directed guidance on when and how to step beyond algorithmic comfort zones for difference and diversity.
    
\textbf{Content Tagging:} Users can optionally label Adventure or Disruption content in the feed to distinguish them from regular recommendations, adding a layer of disclosure and control absent from existing exploration spaces. When enabled, clicking the label opens a brief \textit{“Why You're Seeing This”} explanation for the reasoning and diversity behind this content, with quick access to adjust related settings (see Figure \ref{fig:cd}-3). This design enhances system transparency and streamlines access to backend settings from feeds. We intentionally omit feedback mechanisms here as Adventure and Disruption are meant to preserve surprise and challenge, but feedback could inadvertently suppress diversity and counteract exploratory value by reinforcing preference loops.

\subsubsection{Personal Modes} 
\begin{figure*}
    \centering
    \includegraphics[width=1\linewidth]{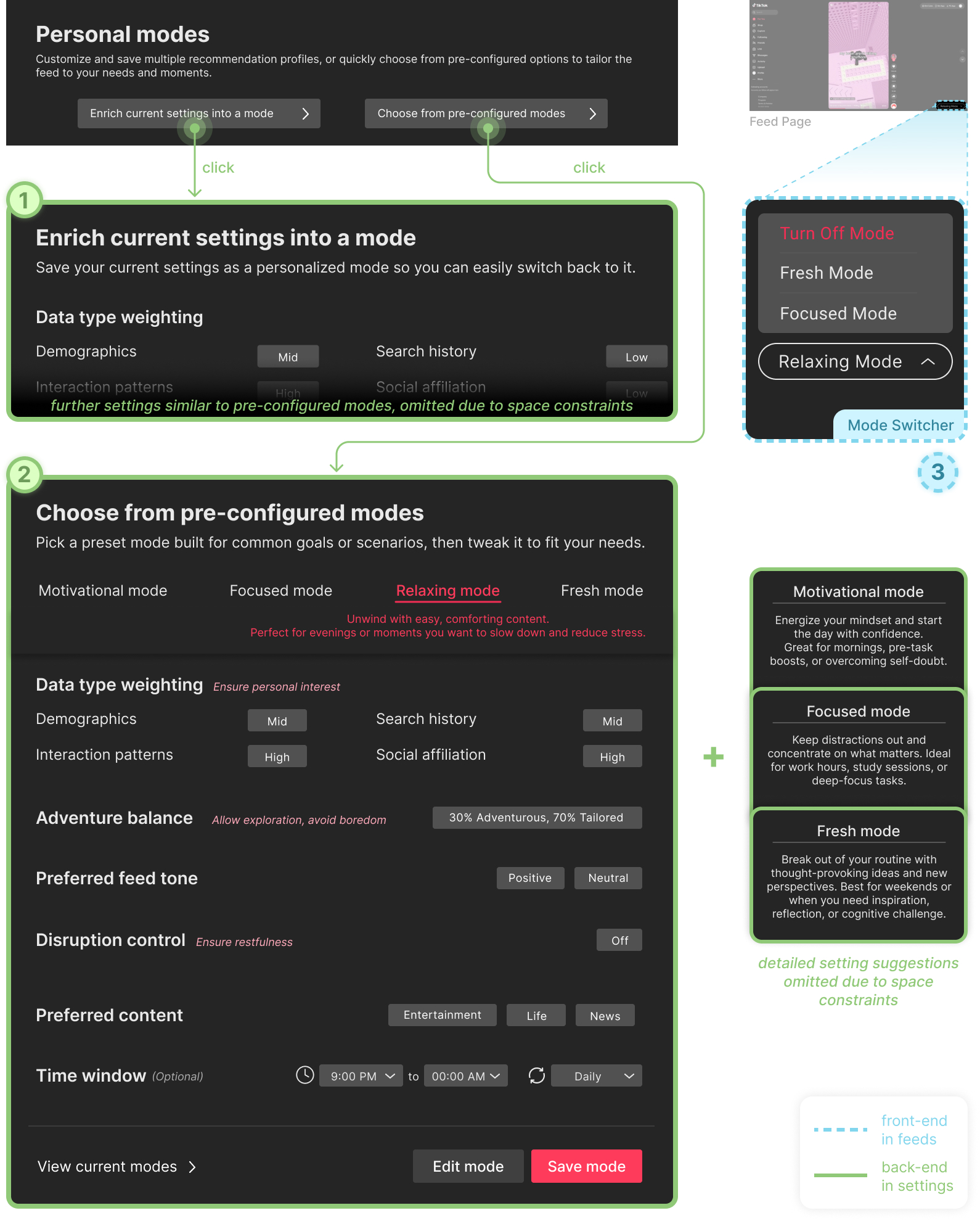}
    \caption{Personal Modes: 1) Enrich Current Settings into a Mode, 2) Choose from Pre-Configured Modes, 3) Mode Switcher.}
    \label{fig:pm}
\end{figure*}

Personal Modes serve as a meta-layer that bundles data and discovery settings into intuitive, scenario-based profiles (see Figure \ref{fig:pm}). Existing RS typically support only a single persistent configuration, which fails to account for users’ shifting moods, goals, or contexts when engaging with media \cite{kompan2013context, kim2024navigating}. Some platforms offer isolated modes, such as TikTok’s “Restricted Mode” or Spotify's mood-based mixes, but they are system-defined with limited user-initiated control.

Inspired by the persona-based interaction feature \cite{wu2025incorporating}, Personal Modes extends this to RS by allowing users to create, save, and switch among multiple configurations tailored to changing needs, such as focusing on work, unwinding in the evening, or seeking novelty. Modes can be created in multiple ways. Users may directly \textbf{enrich their current settings into a mode}, naming and saving it for later use (see Figure \ref{fig:pm}-1). During setup, they can refine the mode by specifying content tone (e.g., positive, neutral, challenging) and focus areas (e.g., work, learning, entertainment). Alternatively, the system offers \textbf{pre-configured modes} as quick-start options and inspiration, including: 1) Motivational Mode - energizing content to boost confidence and prepare for tasks, 2) Focus Mode - streamlined, task-oriented, and distraction-free, 3) Relaxing Mode - comforting, interest-based content with light exploration to avoid boredom, and 4) Fresh Mode - thought-provoking, diverse content to induce reflection and broaden perspectives (see Figure \ref{fig:pm}-2). Beyond manual creation, the system can \textbf{auto-detect browsing patterns} (if any) and prompt users to save them as a mode through system messages. On the front-end feed page, a visible \textbf{Mode Switcher} enables quick transitions between modes or return to the default settings (see Figure \ref{fig:pm}-3). Together, these capabilities support flexible, context-aware control over recommendation behavior, supporting dynamic adaptation as users' needs shift.

\section{USER STUDY}

\subsection{Recruitment \& Participant Overview}
Following IRB approval, participants were recruited via snowball sampling. Recruitment materials described the topic of “personalized recommendation” and the challenge of how such services could better align with user needs. Interested individuals completed a preliminary survey collecting demographic data and their experience with the current RS (see Section \ref{survey}). To capture diverse perspectives, we also asked about their prior exposure to project-related topics (e.g., user agency, AI transparency, privacy).

The final sample included 19 participants (8 males, 11 females; mean age = 28.11, \textit{SD} = 3.96). Twelve were undergraduate to PhD students in fields such as Computer Science, Information Science, UX, and Communication; seven were working professionals in technology, research, or design. Ten participants reported no direct domain experience, while nine had some academic or professional involvement. Because this study involved early-stage provotype probing and sought to generate formative design insights, we selected individuals with general familiarity with algorithmic media and RS. Such participants are better positioned to articulate mental models, compare experiences, and engage critically with speculative features. Although TikTok served as a familiar context and visual scaffold, most participants were active users of TikTok, social media, and other RS (e.g., E-commerce, streaming, news), ensuring sufficient grounding for evaluation. Each participant received a \$15 Amazon gift card as compensation. Table \ref{tab:profile} details their profiles.

\begin{table*}
\caption{Participant Profile: Demographic, Background, and Familiarity with the Project Domain}
\centering
\begin{tabular}{lcclc}
\toprule
\textbf{P-ID}  & \textbf{Gender} & \textbf{Age} & \textbf{Background} & \textbf{Topic Experience} \\
\midrule
P1  & F & 26 & Student in Journalism                   & No  \\
P2  & F & 24 & Data Scientist                          & No \\
P3  & F & 29 & Student in Computer Science             & Yes \\
P4  & M & 26 & Technology Research Assistant           & Yes \\
P5  & F & 34 & UX + Software Development in Finance    & Yes \\
P6  & M & 36 & UX in Information Technology            & No  \\
P7  & F & 26 & Student in Communication                & Yes \\
P8  & F & 33 & Landscape Designer                      & No \\
P9  & F & 30 & Student in Information Science          & No  \\
P10 & M & 23 & Student in UX                           & No \\
P11 & M & 35 & Student in Information Science          & No  \\
P12 & F & 27 & Student in Communication                & Yes \\
P13 & F & 26 & UX Researcher in Digital Health         & Yes \\
P14 & F & 29 & Project Manager in AI                   & Yes \\
P15 & F & 26 & Student in Computer Science             & Yes \\
P16 & M & 24 & Student in Computer Science             & No \\
P17 & M & 24 & Student in Psychology                   & No  \\
P18 & M & 30 & Student in Information Science          & No  \\
P19 & M & 26 & Student in Cognitive Psychology         & Yes \\
\bottomrule
\end{tabular}
\label{tab:profile}
\end{table*}

\subsection{Study Procedure} \label{procedure}

\subsubsection{Preliminary Survey} \label{survey}
Before the study, participants completed a preliminary survey to establish baseline information and contextualize their later feedback. The survey collected demographic data and asked about participants’ prior experience with in-market RS, including the types of systems they used (e.g., E-commerce like Amazon, social media like TikTok, news like BBC, streaming like YouTube, travel like Yelp) and their frequency of use. Participants then reported their baseline perceptions of these systems. Because people typically engage with multiple RS in everyday life, we asked participants to reflect on their general, aggregated impressions of how the RS they use behave. Using items from \cite{pu2011user}, the survey assessed: 1) system transparency (e.g., \textit{“The system makes clear why the content is recommended to me”}), and 2) control (e.g., \textit{“The system allows me to influence the AI-recommended content”}). All items used a 5-point Likert scale (1 = Strongly disagree, 5 = Strongly agree).

\subsubsection{Provotype Interaction and Semi-Structured Interview}
To evaluate the effectiveness and potential impacts of our provotype, we took a qualitative, in-depth approach. This enabled us to gather nuanced reflections, assess the usability and desirability of proposed features, and explore their influence on users' sense of agency. Sessions were conducted remotely via Zoom, with audio recorded, and researchers facilitated interaction and provided support when necessary. Each 75-minute session consisted of three parts:

\textbf{Warm-Up and Baseline Exploration:} The session began with an open discussion of participants’ general impressions of AI recommendations across the platforms they personally use. The facilitator then shared their preliminary survey responses to prompt elaboration on their ratings of clarity and perceived control of current RS, factors they believed influence recommendations, strategies they used to manage them, and any concerns about data usage. This situated participants’ existing mental models and provided reference points to the provotype sessions and subsequent interviews.
    
\textbf{Provotype Walkthrough and Feedback:} Participants were first presented with the provotype feed, where they could browse stimulated posts and access the Content Preference settings to familiarize the context. Newly designed features were then introduced for free exploration. Think-aloud protocols \cite{van1994think} were used to elicit real-time reflections on the goal, usefulness, clarity, and concerns of each feature (e.g., \textit{“How might this feature impact your content exposure?”}). Researchers observed, took notes, and provided clarifications when needed. After this open critique phase, they were asked to prioritize a feature they would keep if only one remained and explain why. This served as a strengths-oriented elicitation strategy to identify which design directions were considered most essential or development-worthy. To mitigate positivity bias, we paired this with open-ended questions about confusion, frustrations, and suggestions for improvement. A brief survey mirroring the baseline questions was then administered to assess perceived transparency and control in our provotype. Drawing on these responses and any shifts from the baseline, a semi-structured interview probed participants’ reasoning behind their ratings (e.g., \textit{“Your control score changed/didn't change, why is that?”}). The interview then broadened to examine how the provotype influenced their understanding of personalization and exposure dynamics (e.g., \textit{“Compared to real-world systems you use, did it change how you think about recommendations? In what way?”)}. Additional prompts explored trust and comfort with data practices relative to existing RS (e.g., \textit{“How about trust? What does the system give to you?”}).
    
\textbf{Co-Design for Adoption and Awareness:} Recognizing the potential cognitive load from diverse settings, the final part invited participants to collaboratively brainstorm strategies for effectively implementing and promoting these features. Discussions centered on increasing adoption, designing approachable interfaces, and avoiding overload (e.g., \textit{“What design would make you more likely to review and understand these features?”}). Broader reflections further explored how to foster awareness of power dynamics and user agency in AI systems to encourage active engagement over passive acceptance (e.g., \textit{“How could we make users more aware of their power in shaping the AI? What would make that feel clear and motivating?”}).

\subsection{Data Analysis}
We adopted a mixed-methods approach to data analysis, combining quantitative comparisons of perceptual measures with qualitative thematic analysis of user insights, allowing us to understand both what changed and why.

Through two targeted surveys administered before and during the study, we quantitatively assessed participants' perceptions of system transparency and control between the current RS and our provotype. Given the modest sample size (\textit{N} = 19), we used Wilcoxon signed-rank tests to compare pre- and post–ratings. The comparative results helped highlight participants' perceptual shifts in transparency and control, and guided our qualitative analysis to explore underlying explanations and experiences in depth.

Qualitatively, all sessions were audio-recorded and transcribed with participant consent. We conducted inductive thematic analysis \cite{braun2024thematic} and cross-referenced transcripts with observational notes. Two researchers independently open-coded five transcripts to generate an initial coding set. Through iterative discussion, overlapping codes were merged and refined to form a shared codebook for the remaining transcripts. Disagreements were resolved through discussion until consensus, with occasional adjudication from another qualitative researcher on the team when needed. The final code structure followed the protocol, resulting in seven themes for general RS experience, eleven for provotype feature reaction, three for provotype-induced changes, and two for adoption/awareness strategies. Subtopics emerged within these themes, such as \textit{“Heightened Concerns from the Visibility of Data”} under \textit{“Shifts in Feeling on Data Usage.”} These qualitative insights complemented the survey findings by explaining why perceptual shifts occurred and surfacing additional system design strategies.

\section{RESULTS}
To address our research goals, we first briefly described participants’ general mental models of existing RS as context. Then, we discussed their specific feedback on provotype features, and how the provotype influenced their perceived transparency, control, trust, and data comfort toward the system. Finally, we present participants’ insights on strategies to encourage feature adoption and to enhance agency awareness more broadly.

\subsection{Mental Models of Current Recommender Systems}

\subsubsection{Transparency Gaps and Reliance on Inference}
Participants widely found the current RS unclear or opaque, with 11 disagreeing in the pre-survey that systems explain recommendations well. Some said the system \textit{“never tells you why”} (P4, P17), while five recalled only generic labels like \textit{"“Recommended for you”} (P8). Several believed this opaqueness was intentional, describing it as a business strategy to \textit{“make profits”} (P11), \textit{“avoid backlash”} (P5), or even to \textit{“make systems seem intelligent and really in tune with you to keep you there”} (P5). Without clarity, 13 participants inferred mechanisms by observing system behavior or by guessing. They cited user behavior (e.g., clicks, watch time, search, 16 participants), demographics (ten participants), and social networks (five participants) as key drivers. Seven even suspected that those precise recommendations came from systems that eavesdropped on their chats with friends, adding to the uncertainty and fear. This limited transparency underpinned participants’ doubts about their ability to influence recommendations.

\subsubsection{Perceived Lack of Control and Desire for Proactive Influence}
Although participants recognized that engagement could shape recommendations, most felt limited control over the systems (12 of 19). Algorithms were viewed as primarily driven by platform priorities and operating \textit{"behind the scenes"} (P1). While 14 participants acknowledged that feedback such as likes/dislikes, blocks, or searches could affect recommendations, they expressed that \textit{"understanding algorithm logic did not translate into feeling in control"} (P11). These actions were perceived as reactive, implicit, and ineffective (P2, P4, P18), with repeated exposure to unwanted or previously dismissed content further undermining the sense of control (P16, P18, P19). Settings offered little relief: few participants explored them, citing \textit{"not expect the settings to help"} (P1, P5, P17), struggled to locate them (P6, P14), or were even unaware of them entirely (P18). Those who accessed found only basic controls (P8, P17, P19) and expressed a desire for more proactive or global settings (P7, P14), such as \textit{"adjusting interests before generating [the recommendations]"} (P7). In practice, users relied on trial-and-error strategies, such as deliberately manipulating engagement to "train" the system (nine participants) or avoiding interactions to prevent leaving traces (four participants). Moreover, perceived ease of control depended heavily on visible cause-and-effect: participants felt empowered when their actions produced immediate, noticeable changes (P3, P7, P17), whereas hidden mechanisms and inconsistent performance often led to frustration (P4, P16).

\subsubsection{Personal Data Concerns}

Participants reported widespread unease over data collection, profiling, and opaque data flows. Many participants feared hidden tracking or surveillance, with six worrying systems \textit{"consistently and secretly watch [them],"} seven concerned about data leakage to third parties for misuse, and three critiquing \textit{"data selling."} Several participants highlighted the risk of recommendations shaping behavior or influencing personal identity without explicit consent (P11, P12, P13), reflecting broader concerns about covert manipulation. While two participants accepted data collection as an inevitable trade-off for personalization (P10, P17), most still sought clearer disclosure and filtering options to understand what data is used and how.

\subsubsection{Reflections on Current Personalization} Participants expressed mixed experiences with their existing personalized feeds. Some valued the usefulness in \textit{“discovering unknown needs”} (4 participants) or \textit{“adding search sources”} on their behalf (P14). Others noted mismatches between recommendations and their actual interests, calling them \textit{“sometimes irrelevant”} (P10) or \textit{“random in accuracy”} (P14, P18). However, eight participants welcomed irrelevance. They expressed frustration with narrow and repetitive feeds, describing them as \textit{“too similar”} and \textit{“stuck,”} and felt annoyed or anxious when the system kept pushing persistent themes. They explicitly linked this to filter-bubble effects that could \textit{“create bias”} (P3) or \textit{“narrow down the world”} (P5), and wished they could \textit{“opt-out,” “reset,”} or \textit{“manually curate”} what they see (P14, P16). Four participants depicted concrete control over the degree of personalization, such as \textit{“bypassing past preferences and see non-personalized content,”} and envision adjustable balances like \textit{“3/4 familiar, 1/4 different”} to break from sameness and gain enlightenment (P3, P14).

\subsection{Feature-Level Reactions to the Provotype} \label{feature}
\begin{figure*}
    \centering
    \includegraphics[width=1\linewidth]{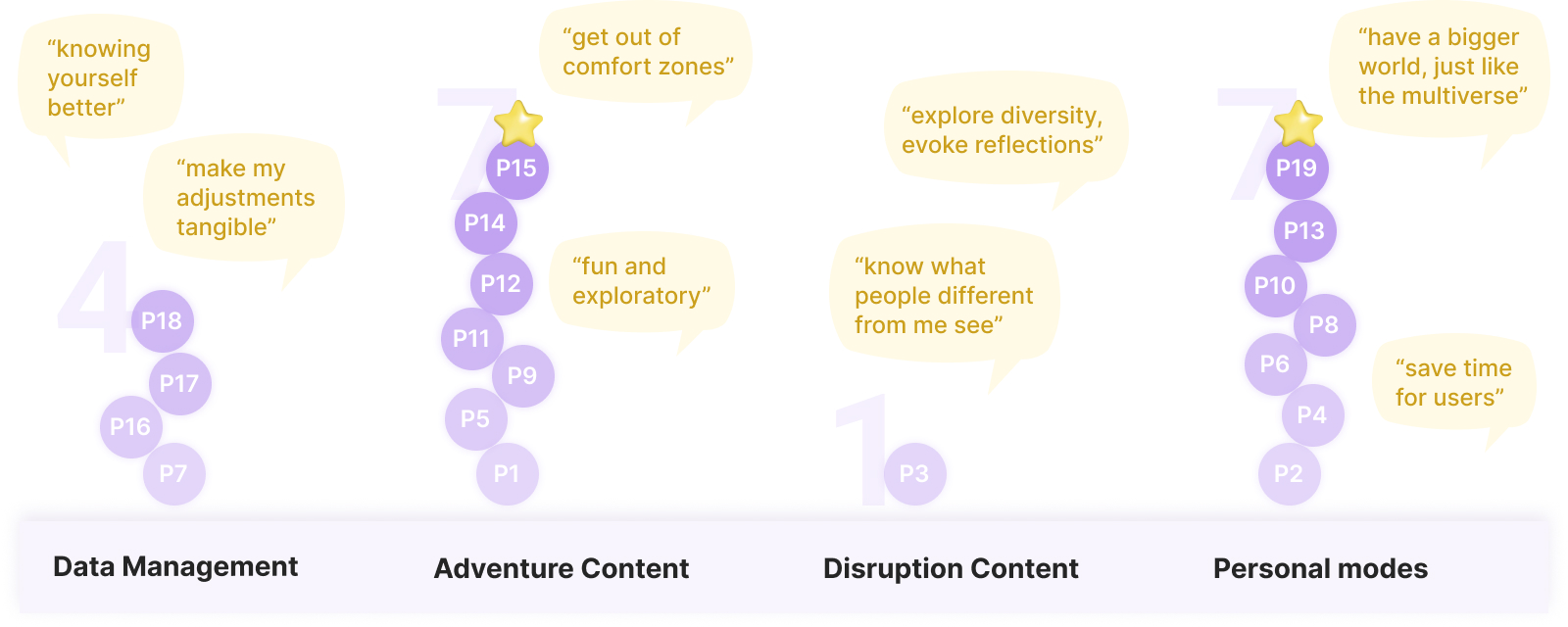}
    \caption{Participants' Votes for Favorite Provotype Features with Representative Reasons.}
    \label{fig:keep}
\end{figure*}

This subsection reports participants’ immediate reactions to the design of provotype features during the walkthrough, focusing on perceived usefulness, clarity, and concerns. These reactions reflect localized judgments about what each feature affords, how intuitive it felt, and where design improvements could be made. Participants also described how specific controls would change their personalized feeds. To understand feature desirability more directly, Figure \ref{fig:keep} shows participants’ votes for the most compelling one with representative reasons. \textbf{Adventure Content} and \textbf{Personal Modes} were most frequently chosen (seven votes each), suggesting strong promise for initial deployment in practice.

\subsubsection{Data Management}
Participants generally welcomed the \textbf{Data Type Weighting} as empowering. P1 indicated that existing in-market settings are \textit{"too simplistic"} that based on coarse, binary cues, but now \textit{"it can have a specific weight."} Seven supplemented that this adjustment let them suppress unwanted signals and emphasize preferred ones, particularly in contextual adjustments (e.g., increasing search history for focused tasks, decreasing social affiliation to reduce friends’ impacts). This feature was valued as a personalization tool that encourages experimentation and active \textit{"training"} of the algorithm, rather than purely a privacy measure. However, P10 and P13 expressed concerns and discomfort about \textit{"feeling objectified"} by granular distinctions, and P13 raised the need for rethinking the categories to be \textit{"more natural to users,"} such as combining search history and interaction patterns since they are \textit{"highly related in practice."} Some participants desired even finer controls beyond our version to use numeric scales (0-100) instead of intervals (P11, P18). P10 also desired clearer explanations for why each data type affects recommendations. 

Relevant to data weighting, the \textbf{Data Usage \% Indicator} was praised for boosting transparency and making their adjustments tangible (four participants). P3 expressed that those indicators \textit{"are ways to understand myself and refresh my behaviors and preferences,"} especially when encountering confusing recommendations. Some participants said they would actively check it (P16) and suggested including more visual aids like bar charts to enhance the clarity of percentages (P2, P11). Yet, interest in this feature was split: some doubted its notability and appeal, as it may \textit{"feel troublesome"} by adding cognitive load beyond content browsing (P9). Lighter alternatives to increase exposure with less manual effort were proposed, such as pop-up breakdowns following user interactions (e.g., after they "like” a content) (P14) or periodic summaries (monthly/annual) that would make insights more meaningful and shareable (P2).

The \textbf{Data Aging} setting was highly favored, seeing as intuitive and aligned with users' changing interests. During the walkthrough, participants gave immediate reactions as \textit{"favorite in Data Management... more controllable than others"} (P4) and \textit{"easiest thing I can understand"} (P13). Its time-based framing resonated, as P7 said, \textit{"I wouldn’t want content from three months ago."} Participants valued its potential to correct algorithmic overreactions to temporary actions, such as \textit{"undervaluing occasional clicks or likes that do not reflect trust interests"} (P3). It could also help explore older content and revisit past interests when needed (P9, P17). However, some preferred this function for situational use rather than daily, suggesting changing it to an on/off toggle for convenience (P10)

\textbf{Data Diary} prompted substantial debate during the walkthrough. On one hand, five participants found it particularly transparent and informative to \textit{“know yourself better with quantified effects”} (P11). They framed it as a personal journey for reviewing and reflecting on interests or behaviors over time to help guide recommendation adjustments (P16, P18), akin to self-tracking (P1, P18) or financial spending apps (P14). On the other hand, participants reported complex emotions when they saw those records. Five felt discomfort as being \textit{“watched”} or noted \textit{“rebellious vigilance,”} while others reconciled this tension: \textit{“at least you know”} (P16, P18) or \textit{“it’s normal, just the system decides whether to tell us”} (P3). The most critique raised by five participants was the lack of actionable controls, leading to passivity and limited utility. They requested options, such as editing/deleting off records or stopping specific entries to prevent unwanted influence. P17 explicitly linked such modifying power to a higher likelihood of use. Other suggestions included more structured summaries (weekly/monthly) instead of exhaustive logs (P12, P17, P19), and the use of visualizations or hierarchical views to improve comprehension (P2, P12).

\subsubsection{Content Discovery} Nine participants strongly appreciated the \textbf{Adventure Content} as refreshing, exciting, and fun, offering a way to counter monotony and narrow exposure by helping users \textit{“get out of comfort zones.”} They envisioned use cases like boredom relief (P9, P12), news discovery (P10), or even dating apps (P19). P3 and P14 visibly lit up and gestured excitedly upon discovering this feature, immediately recognizing it as what they had long wanted.

\textbf{Disruption Content} was similarly valued for breaking filter bubbles and exploring diverse perspectives (four participants). Compared to the more random Adventure, Disruption was praised for opposite-but-relevant content, especially to \textit{“know what people different from me see”} and prompt reflection (P8, P11). However, unlike the enthusiasm for Adventure, participants were more hesitant about Disruption, seeing it as more situational and better suited to niche users like researchers or explorers who are not only seeking relaxation or default browsing (P2, P3, P16). In this context, the ability to limit topics was particularly valued by ordinary users, as P2 indicated, \textit{"I would never want to be disrupted for some topics, but honestly, it'd be cool to see how people in other places for culture or industry. It depends on context."} Participants also recommended renaming it to “Exploration” or “Contrast” to sound less negative and encourage acceptance rather than evoking fear (P10, P14).

\textbf{Time-Based Scheduling} elicited mixed reactions: P14 valued Adventure/Disruption Time as \textit{"completely in line with needs"} (e.g., before or after work), while five others resisted rigid timing: as P3 explained, \textit{"I don’t follow a fixed schedule to use platforms. I think most people use them randomly, so this feature may not be very useful."} Instead, they suggested quick toggles in upfront feeds to \textit{"make it timely in a more accessible place, similar to 'one-click protection.' If you see too many posts you don’t like, you can protect at once"} (P3). Alternatively, occasional pop-ups like \textit{“Are you satisfied with current recommendations?”} can also help remind of the settings (P12).

Similar to Data Usage \% Indicator, \textbf{Content Tagging} was appreciated for its transparency and connection to participants' actions, helping them understand why certain content appeared, prompting self-reflection, and informing the adjustments on content balance (e.g., increasing certain content types if found interesting) (P2, P3).

Overall, six participants valued control over content discovery. They enjoyed the contrast between controllable randomness (Adventure) and intentional diversity (Disruption). For higher simplicity, some even suggested merging the two into a single slider with opposing ends (left: Adventure, right: Disruption) (P7, P14).

\subsubsection{Personal Modes}
Participants widely found Personal Modes appealing and innovative, appreciating how they simplified recommendation control and easily adapted feeds to different moods, tasks, or times of day to \textit{"save time for users"} (12 participants). Some likened it to \textit{"creating profiles"} or a \textit{"multiverse"} to \textit{"having a bigger world"} (P5, P11). This feature offered flexibility and agency without requiring multiple accounts or constant fine-tuning (P16, P18, P19).

However, concerns centered on complexity, learning cost, and potential over-control (P5, P13, P15). If modes were too time-consuming, hidden, or difficult to understand, participants (especially casual users for simple, playful experiences) might avoid exploring them (six participants). To reduce complexity and learning costs, most participants (seven) viewed pre-configured templates as easier and more approachable to start with than building from scratch. These presets were also trusted as deliberate, logical choices by the platform and should be ideal starting points for personalization (P8, P9). Suggestions for reducing cognitive load included using visual metaphors (e.g., avatars, GIFs, emojis) to illustrate modes (P4, P5, P14) and surfacing modes during onboarding with intuitive explanations and quick-start settings (four participants). P14 also desired for content-based mode creation: allowing users to generate a mode directly from a piece of content, for example, \textit{"pushing similar items for the next three hours."}

Potential counter-effects were also considered: inappropriate use of modes could reinforce filter bubbles if users stick to a single preferred mode (P15), and interactions in one mode might unintentionally affect others, causing confusion or frustration (P3, P17). In response, participants emphasized data separation that allows users to choose whether to save or discard mode-specific data to avoid cross-mode contamination (P17).

\subsection{Provotype-Driven Shifts in Perceived Transparency and Control}
Beyond feature-specific design reactions, we also examined how the provotype as a whole shaped participants’ broader perceptions of transparency and control using both a short survey and a follow-up interview. The survey establishes the measurable shift, and the thematic data explains the underlying reasons.

Participants showed great improvements in perceived transparency and control when comparing their experiences with existing RS to the provotype. As shown in Table \ref{tab:survey}, Wilcoxon signed-rank tests revealed substantial increases in perceived transparency (from 2.95 ± 0.74 to 4.63 ± 0.40, \textit{V} = 190, \textit{p} < .001) and control (from 2.77 ± 0.91 to 4.51 ± 0.32, \textit{V} = 171, \textit{p} < .001). Rank-biserial effect sizes were large for both (transparency: \textit{r} = 0.88, control: \textit{r} = 0.87), indicating strong practical significance. These results suggest that our design interventions were highly effective in enhancing participants’ sense of agency in RS (reflected in transparency and control).

\begin{table*}
\centering
\caption{Comparative Statistics for Transparency and Control between Current Recommender Systems and Our Provotype.}
\begin{tabular}{lccccc}
\toprule
\textbf{Metric} &
\textbf{M $\pm$ SD (Current)} &
\textbf{M $\pm$ SD (Provotype)} &
\textbf{V} &
\textbf{p-value} &
\textbf{Effect Size ($r$)} \\
\midrule
System Transparency & 2.95 $\pm$ 0.74 & 4.63 $\pm$ 0.40 & 190 & <.001 & 0.88 \\
Perceived Control   & 2.77 $\pm$ 0.91 & 4.51 $\pm$ 0.32 & 171 & <.001 & 0.87 \\
\bottomrule
\end{tabular}
\label{tab:survey}
\end{table*}

Interview data illuminate what drove these improvements. Across sessions, participants reported higher awareness of recommendation mechanisms and stronger feelings of agency, describing the provotype as reshaping their view of RS from opaque forces into visible, controllable, user-driven tools. Eight participants explained that the provotype replaced guesswork with a more structured understanding of why recommendations appeared and how their data was used. As P17 summarized, \textit{“it turns scattered knowledge into a structured view with subfields and modules.”} Even small indicators, such as those disclosing data usage or content type, helped participants link their own actions to outcomes and signaled that they could shape their feeds (P4, P8, P18). This visibility made recommendations feel less arbitrary and encouraged more willingness to engage, while also helping participants realize \textit{“how the settings can be used”} (P18), in turn, fostering greater agency and willingness to explore control options (P8, P16).

Alongside increased awareness came a stronger sense of control. Proactive features like feed customization and switchable modes made the system feel \textit{“more user-centric”} (P4, P6) rather than \textit{“controlling [users]”} (P9) or \textit{“applying the same algorithms to everyone”} (P8). Compared with real-world apps that felt \textit{“one-way and general,”} the provotype was seen as more innovative and granular (P19), as well as adaptive to build their own platform (P1, P4). P14 further described it as \textit{“a complete process of receiving, understanding, and controlling recommendations”}: first noticing indicators and understanding why content was recommended, then gradually being “educated” by repeated exposure, then having options to fine-tune recommendations, and finally using a portable Mode Switcher to simply align feeds with moment-to-moment needs.

\subsection{Provotype-Driven Shifts in Trust and Data Comfort}
Beyond transparency and control, participants reflected on how the provotype shaped their broader feelings of trust in RS, while also exposing limits to control, nuanced privacy concerns, and mixed emotions about data visibility. These findings highlight the complex interplay among transparency, agency, and trust, showing both the benefits of clarity and the boundaries of user influence.

\subsubsection{Trust Built on Transparency and Control}
The newfound clarity and control shaped participants' evaluations of system trustworthiness. Seeing how recommendations were generated, coupled with meaningful levers of influence, fostered feelings of safety and empowerment (nine participants). For instance, P14 and P19 described having \textit{"large decision-making power,"} while P16 perceived that he \textit{"had a higher role toward the system and didn't have to depend on or hope for certain outcomes."} Importantly, this sense of safety was not always about absolute guarantees, but rather about at least having visible options. Participants noted that, even if they did not use the features or were unsure about their real effects, simply knowing they existed contributed to greater trust and confidence in the platform (P8, P14). Consequently, six participants explicitly favored this provotype over existing platforms because \textit{"it cares more about users"} (P18) and is run by a company \textit{"more responsible and aware of people's preferences"} (P15, P16), resulting in noticeably higher willingness to engage with such platforms.

\subsubsection{Hesitation on Full Trust}
Still, trust could be conditional and nuanced. Despite positive shifts, participants hesitated to claim perfect control or trust. Three emphasized that such completeness is unattainable due to the inherent complexity and opacity of RS: P8 said, \textit{"there must still be many algorithms behind it, you cannot overcome AI’s power,"} while P18 added, \textit{"it would still be a black box because we don’t and can’t know the inner workings."} Others pointed to external limitations, such as third-party influence or advertising, that would continue to shape recommendations outside users' control (P19). Additionally, participants stressed that trust ultimately depends on \textit{"how well it works in practice,"} with accuracy, stability, and evidence of responsible data use all determining whether trust could be sustained (P11, P14, P15). Especially, presenting greater empowerment also raised user expectations: \textit{"if platforms promise clarity and control, they must perform well, correspondingly, and consistently, otherwise frustration will deepen"} (P3).

\subsubsection{Data Concerns and Emotional Ambivalence}
Beyond those structural limits of trust, participants drew distinctions between trusting the algorithm itself and trusting how their data is handled. Comfort of data usage largely improved when the provotype demystified the "black box" of recommendations (P13, P18, P19), but data concerns were not fully alleviated because \textit{"it still collects and tracks my data"} (P10, P12, P16). For some, reassurance seemed superficial: P15 uniquely reported that abundant controls made her \textit{"less think about data brokerage or selling but just how to tweak my feeds,"} suggesting that detailed options may overshadow, but not resolve, underlying data exploitation concerns.

Emotionally, participants expressed ambivalence. The provotype's transparency made the system feel clearer, safer, and more empowering (P7, P14, P18), but explicitly seeing the extent of personal data use and realizing how much platforms know about them also heightened privacy concerns (P8, P14). Yet, although unsettling, participants preferred clarity over ignorance, as P18 observed, \textit{"even if it is scary, I'd rather not be in the dark."}

\subsubsection{Nuanced Value of Transparency}
Participants consistently emphasized that transparency, while valuable, was insufficient on its own to build trust. Some described it as \textit{"a minimum requirement rather than a guarantee of trust,"} with disclosure improving efficiency and clarity but not fundamentally altering confidence in the system (P2, P3, P12). In some cases, detailed metrics may even spark skepticism, requiring users to put in extra mental effort to question the authenticity of the displayed information (P15).

Three scenarios were commonly compared: no disclosure, disclosure without control, and disclosure with control. Participants agreed that the worst case was being unaware of any system workings, while merely revealing information created what P18 called an \textit{"illusion of control"} that offered reassurance that the platform was not hiding things, yet \textit{"still frustrating when powerless to act."} Others echoed this as \textit{"static, non-editable disclosures may amplify the sense of surveillance"} (P4). By contrast, four participants highlighted that coupling transparency with actionable features unleashed its power, allowing them to feel both informed and empowered. 

\subsection{Insights for Encouraging Feature Adoption}
Participants widely recognized the value of provotype features for personalization and agency safeguard. However, six participants stressed that the premise of benefiting from these features is awareness, as P15 noted, \textit{"they need to notice it exists before planning to adopt it."} The next factor would be usability and ease of use, with participants cautioning that perceived complexity, technical jargon, and information overload could discourage regular interaction (as discussed in Section \ref{feature}). In response, participants co-designed strategies to make features easier to discover, simpler to understand, and more inviting to use. While these discussions were grounded in our provotype, the resulting strategies extend beyond this context and offer broader implications for diverse platforms and scenarios.

\subsubsection{Onboarding and Embedded Guidance} \label{adoption}
Participants identified education and timely guidance as the first step toward adoption. Six suggested tutorials or onboarding for initial awareness, and eight highlighted that guidance should also appear within everyday browsing for timely help rather than buried in menus. Indicators and Tagging features in the provotype were cited as effective examples of integrated guidance, with P15 marking, \textit{“providing shortcuts to connect the front and back is a very easy way to access relevant settings when the need arises.”} Four participants proposed contextual prompts as additional shortcuts, such as system messages that appear periodically or in response to user actions: \textit{“Do you want some adventure?”} (P4) or \textit{“If you need [certain things], set it up here!”} (P5). These in-flow cues align more with users’ natural behavior and intent (i.e., browsing content) than requiring deliberate detours to settings. For all those efforts, P13 specially noted, \textit{“even if users do not actually use them, just seeing these settings is already an education”} and can motivate later exploration.
    
\subsubsection{Layered and Adaptive Designs}
Once aware, users need to understand how to engage without being overwhelmed. To lower this barrier, seven participants suggested layered controls: a basic, simplified version paired with an “advanced settings” for deeper customization, described as \textit{“a process of acceptance”} (P14). This also accounted for different user types, recognizing that casual and power users would have different needs (P4, P7, P11). Building on Personal Modes, six participants recommended \textit{“just packaging everything into several predefined modes”} to simplify decision-making, as \textit{“one-click to switch mode is much simpler than adjusting detailed settings”} (P3, P4), but enthusiasts could still access a \textit{“see details”} option for finer adjustments.

Layers could also be extended across devices. As P18 proposed, \textit{“after learning the concepts, smaller devices could have a streamlined version, such as only keeping modes on mobile, and users return to the desktop for complete settings,”} supporting progressive and context-appropriate engagement.
    
\subsubsection{Goal-Oriented Framing and Engaging Process}
Beyond awareness and comprehension, willingness to use such controls could depend on how appealing they feel. Participants suggested framing them as exciting, goal-driven tools rather than dry settings. For example, instead of segmented elements like drop-down menus (i.e., Mode Switcher) or proportion sliders (i.e., Adventure/Disruption), these modes could directly appear as front-end navigation tabs labeled “For You / Explore / Adventure” to trigger curiosity and invite interaction (P3, P5).

How control is framed also matters. Rather than presenting them as plain settings, participants suggested resonating with users’ everyday motivations, such as productivity, creativity, or enjoyment (P7, P8), and should be context-specific. For example, entertainment systems could \textit{“present them as playful and part of the fun rather than a chore,”} while productivity tools might frame it as a way to \textit{“streamline tasks and save time”} (P8).
    
To sustain adoption over time, participants provided strategies to keep the experience engaging. Gamification elements, such as points or rewards for using these features, could encourage early and continued use (P1). Visual and content cues, including lively tones (P14), concise and bold wording (P18), clean design (P10), and playful videos or cartoons (P16), could all help reduce perceived effort and maintain attention.

\subsection{Insights for Long-Term Agency Awareness and User Empowerment}
Beyond specific features, we broadened the conversation to how people often engage with AI systems passively without realizing their power to influence outcomes. Four participants attributed this to the perception of AI as an inscrutable \textit{“black box”} that reduces curiosity and discourages user control. To counter this, participants proposed ways to cultivate long-term mindsets that help people better recognize and care about their agency and power. 

\subsubsection{Building Lasting Habits Through Visibility and Literacy}
Participants emphasized that users are more likely to feel empowered when systems consistently signal their ability to influence outcomes (P6, P16). This visibility goes beyond feature-level initiatives for discovering settings, reflecting a broader stance in which systems continuously reinforce the user’s role as the ultimate decision maker. Without such affordances, P14 warned, \textit{“it is actually the platforms that 'domesticate' users into passive habits,”} any initial sense of agency can fade when reinforcement diminishes over time. Complementing this view, four participants recommended broader educational efforts, such as societal discourse, community learning, or formal AI literacy campaigns, to help normalize understanding of AI decision processes and the notion of user agency. P18 further suggested introducing such literacy early, so younger users can develop a baseline sense before passivity takes hold. These reflections stress the importance of sustained reinforcement and deeper conceptual understanding for building durable agency habits.

\subsubsection{Responsibility on Designers and Regulators, Not Users}
Notably, five participants argued that agency awareness should not rest solely on users. Instead, \textit{“decision makers at higher levels are the ones who should take the initiative”} (P5, P19). Ethical product designers and system developers were described as \textit{“essentially responsible for building these [features]”} (P13) and for \textit{“reminding users of their ability”} (P12). Participants even elevated it beyond the business level to the societal level, suggesting that regulators should apply external pressure (P5, P14, P19): privacy laws and cookie regulations were cited as examples of how regulation can enforce visibility and gradually normalize awareness, even if adoption remains slow. P5 specifically likened platforms and policymakers to \textit{“adults”} tasked with guiding \textit{“children”} (users) toward safer experiences and \textit{“deciding what to show,”} rather than expecting users to carry the burden of understanding complex systems. Meaningful agency, therefore, requires shared responsibility beyond user action.

\section{DISCUSSION}
This study illustrates how actively involving users can inform the design of recommender systems. By exploring participants’ mental models, evaluating a provotype, and co-designing strategies, we gained insights into how transparency, control, and agency influence user experiences. While some findings are specific to our provotype, many extend more broadly to other personalized systems, offering both immediate design guidance and deeper theoretical insight. Practically, we highlight ways to encourage adoption, strengthen user empowerment, and balance transparency with trust. Theoretically, we also surface enduring dilemmas and socio-technical tensions, revealing the challenges of reconciling user agency with system design and business imperatives.

\subsection{Summary of Findings}
Through provotype interaction and interviews, we found that the proposed features targeting data management, content discovery, and contextual recommendation can meaningfully surface hidden processes and create actionable points of influence in RS. The provotype prompted participants to reframe RS from opaque, unilateral systems into adjustable tools that better align with their goals and preferences. These design interventions led to notable perceptual and attitudinal shifts: participants reported heightened perceptions of transparency and control, stronger trust, and greater engagement compared to existing RS. Such feedback indicates promising directions for user-facing features to address information and power asymmetries by reducing opacity, distributing agency, and alleviating the helplessness often experienced in AI-driven RS.

Ambivalence was also revealed: greater data disclosure raised privacy concerns, excessive options risked cognitive overload, and trust remained conditional given the inherent complexity of RS. This underscores that supporting agency requires careful calibration among clarity, control, and user comfort, rather than simply adding more controls.

Participants also articulated how these controls would reshape their feeds. Adjusting data types and temporal scope reduces moments of irrelevance and allows the feed to “drift” with their evolving interests rather than lag behind. Controlling the balance between familiar and unfamiliar materials offers structured ways to break from repetition and predictable content for novelty. Personal Modes were seen as enabling shifts between different “states” of personalization, reflecting their dynamic and situational nature.

To make features effective in practice, participants proposed strategies such as raising awareness through onboarding, embedding subtle affordances in workflows, layering controls for progressive learning, and framing features around concrete goals, all of which could remind their sense of agency and facilitate adoption. Especially, agency responsibility was seen as shared: designers, developers, and regulators, not just users, play a role in enabling meaningful empowerment. Together, these insights provide a roadmap for informing RS design and empowering agency while acknowledging practical and socio-technical limitations.

\subsection{Design Takeaways for AI Recommender Systems}
\begin{figure*}
    \centering
    \includegraphics[width=1\linewidth]{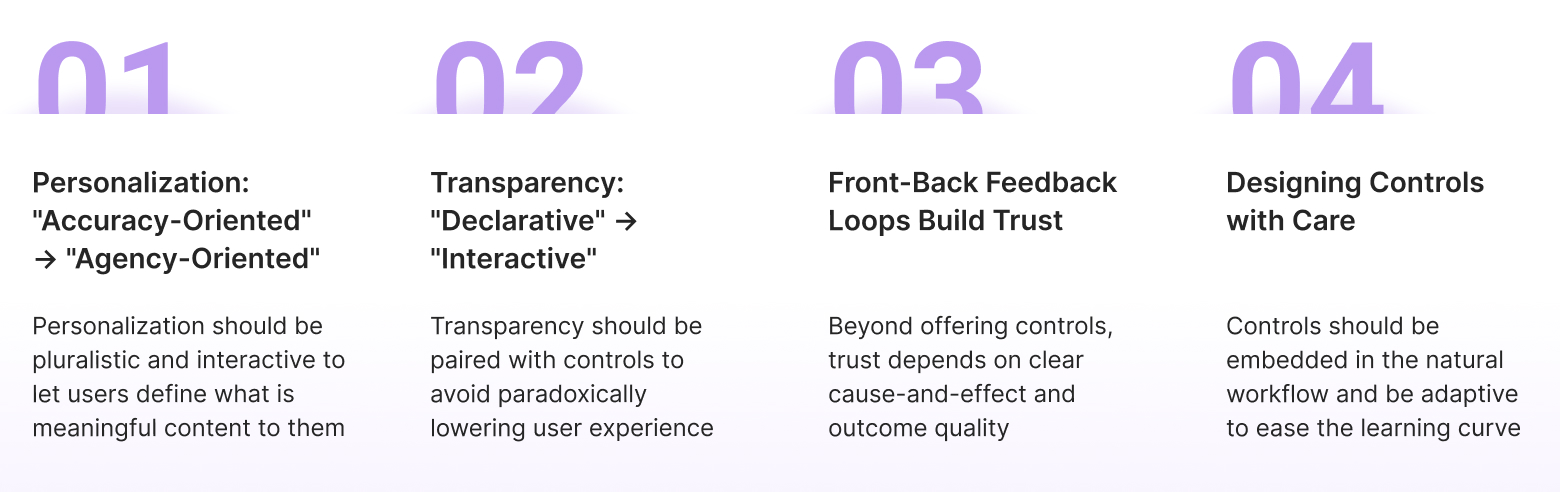}
    \caption{Design Takeaways for AI Recommender Systems.}
    \label{fig:takeaway}
\end{figure*}

As a proof-of-concept designed to enhance users' sense of agency, our provotype yielded promising results. The findings point to concrete design opportunities for both RS and broader AI, showing how agency can be supported in ways that resonate with users’ natural behaviors and values. Figure \ref{fig:takeaway} summarizes our design takeaways.

\subsubsection{From "Accuracy-Oriented" to "Agency-Oriented": Rethinking Personalization}
A key motivation for our work was the recognition that accuracy-focused recommendations can narrow content exposure and reinforce filter bubbles \cite{pariser2011filter, aridor2020deconstructing}. We explored how giving users direct control could offer a complementary path, allowing them to proactively introduce randomness or divergence. Consistent with prior work \cite{wang2022user, cheng2017learning}, our study shows that users already value and desire such options even before encountering our provotype, suggesting that accuracy and relevance should no longer be the sole goal for RS. Participants’ reflections on envisioned feeds also reveal that users do not view personalization as static but flexible and adaptive to their shifting intentions and contexts, supporting prior work on contextual recommendations \cite{hariri2014context}. Different users, or the same user at different times, may seek different personalization patterns.

Unlike backend algorithm refinements to introduce diversity \cite{wilhelm2018practical, adomavicius2009toward} or context awareness \cite{santana2020contextual}, our approach evaluates external, user-facing control designs to recalibrate data influences and loosen or tighten personalization as needed. We argue that future systems must move beyond forced personalization toward a more \textbf{\textit{pluralistic and interactive paradigm}} where users can define what counts as meaningful content in context, rather than remaining passive recipients, even when the system already introduces backend variations. This reframing positions personalization as one option within a broader spectrum of user-system interactions, with controls designed to invite adaptation, exploration, and responsiveness to shifting needs while foregrounding user agency.

\subsubsection{"Interactive", not just "Declarative": Rethinking Transparency}
Participants made clear that, while transparency remains vital in ethical AI design, static explanations of algorithms alone often fall short, or even frustrate, when users cannot act on the information provided. Prior work shows that explanations can boost awareness of algorithmic processes \cite{rader2018explanations}, but awareness (the perceptual agency) does not translate into true agency. As exemplified by the Data Diary, users became acutely aware of system behavior and data collection, but still felt powerless when unable to intervene. This resembles cognitive dissonance theory \cite{harmon2019introduction}, where being informed yet unable to act creates discomfort and lowers perceived agency. Building on the human–AI synergy framework \cite{sundar2020rise} and co-design insights stressing the pairing of transparency and user interventions during algorithm interactions \cite{storms2022transparency}, we argue for reframing transparency \textbf{\textit{from a declarative model}} (where systems merely tell users how they work) \textbf{\textit{toward an interactive model}} (where users learn by doing and observing) to avoid paradoxically diminishing user experience. This extends research on \textit{"interactive transparency,"} where systems \textit{"afford users the option to provide feedback"} \cite{molina2022ai}, by providing more novel approaches that give users a more active role in managing recommendations, thereby enhancing trust and informing future RS design.

\subsubsection{Front-Back Feedback Loops Build Trust}
Agency does not stop at providing controls (the behavioral agency). Even with abundant options, participants stressed the importance of linking back-end actions to front-end outcomes to perceive tangible effects. When adjustments led to clear, visible changes in recommendations, such as through the Data Usage \% Indicator, participants would feel empowered, whereas opaque or delayed results may raise doubts about system legitimacy \cite{herder7feedback, amershi2014power}. Performance also mattered: poor recommendation quality could undermine engagement and hinder participants from achieving full trust or agency. This aligns with prior work showing that perceptions of algorithmic fairness and trust depend heavily on observed outcomes \cite{wang2020factors} and that feedback loops sustain user engagement and confidence in AI systems \cite{krauth2025breaking, zhao2018explicit}. Trust, therefore, is not a fixed property granted by controls but an ongoing state contingent on visible cause-and-effect feedback and system performance. Our findings suggest that effective agency combines both doing (the ability to act) and seeing (evidence that actions matter). Rather than treating control as an isolated feature, designers should integrate performance, responsiveness, and usability to support a robust sense of agency.

\subsubsection{Designing Controls with Care: Embedded, Balanced, and Adaptive}
Many participants admitted they rarely explored deep settings menus, instead preferring lightweight, in-context adjustments. This suggests that controls or explanations should not only be hidden at the periphery but surfaced directly into daily interactions, such as in feeds or via contextual hints. Such embedded cues not only reduce friction and enhance usability \cite{dodeja2024towards, amershi2014power} but also act as educational affordances, helping users stay aware and engaged \cite{thompson2013s, molina2022ai}. In contrast to prior approaches that focused on settings \cite{wang2024chaitok, bemmann2024privacy}, our study advocates a design paradigm that embeds agency into users' natural workflow to align with their original intent and typical behaviors on the platform.

Yet, the balance is delicate. Our study reveals the challenge of operating agency without overwhelming users with complexity. Simply adding more control options does not necessarily increase users' perceived control, but can plateau or even backfire (overload, frustration) to reduce adoption \cite{bemmann2024privacy, storms2022transparency}. While transparency must go beyond simply exposing system processes to influence outcomes, the collaborative relationship should be made meaningful and effective by presenting clear and intuitive control mechanisms without requiring extensive learning burdens. Moreover, while some participants welcomed detailed indicators and granular settings, others voiced concerns and found it unnecessary, indicating that personality also influences control preferences \cite{harper2015putting, rook2020engagement}. Hence, adaptive or layered systems that tailor mechanisms to individual needs can ease the learning curve and maintain engagement. This aligns with the principle of progressive disclosure in interface design \cite{tidwell2010designing} and customizable recommender interfaces \cite{knijnenburg2012explaining, zhang2019proactive}, which manage complexity and improve accessibility by gradually introducing controls to match user expertise.

\subsection{Tensions and Dilemmas in Decision Making during Recommender System Development} \label{tension}
\begin{figure*}
    \centering
    \includegraphics[width=1\linewidth]{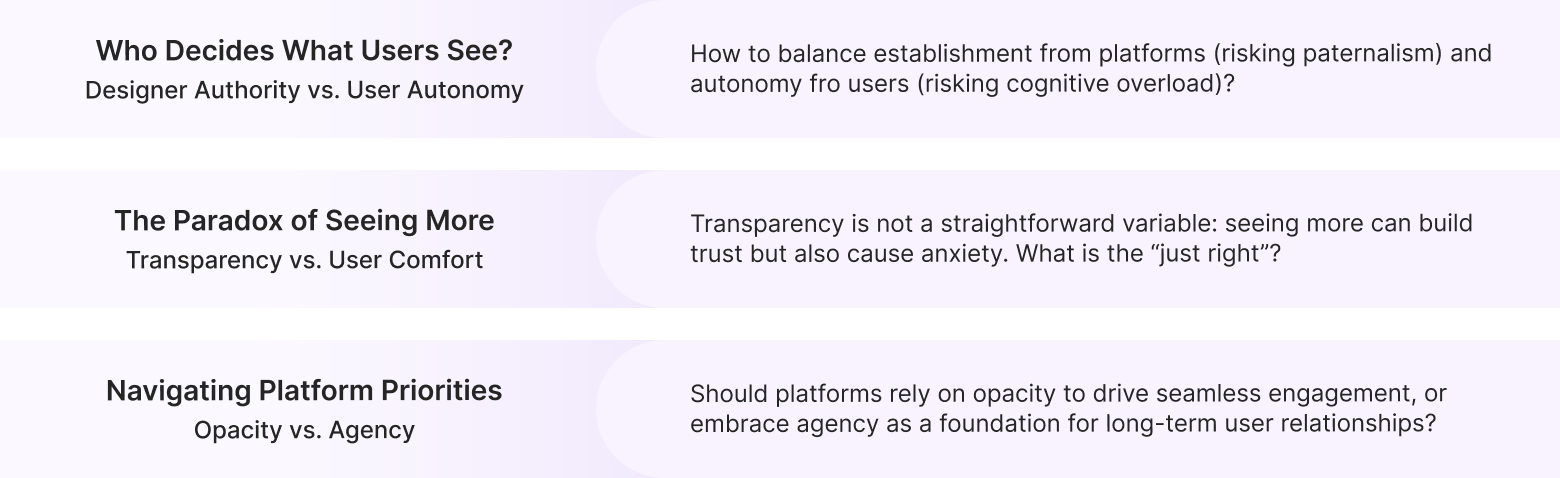}
    \caption{Tensions and Dilemmas in Decision-Making during the System Development}
    \label{fig:tension}
\end{figure*}

Our study also surfaced several tensions that extend beyond immediate design choices, highlighting broader socio-technical dilemmas in developing RS. Derived from the design takeaways, these tensions reflect trade-offs between user empowerment, system performance, and business priorities, showing that designing for agency is rarely straightforward. Instead of prescribing solutions, examining these dilemmas helps illuminate the philosophical and strategic questions underlying personalization, transparency, and control, offering guidance for both research and practice. Figure \ref{fig:tension} illustrates the discovered tensions for RS development.

\subsubsection{Who Decides What Users See? Designer Authority vs. User Autonomy}
The challenge of balancing controls raises a broader dilemma: who decides what users see, and how should such decisions be made? Users have long criticized RS for making choices on their behalf. When developers fully determine what to present, user agency is clearly missing, as many scholars point out \cite{sundar2020rise, wu2025negotiating}. However, granting users full control is not a simple fix either. While seemingly empowering, the more aspects users can control, the more complex and overwhelming the framework becomes, which could worsen the service \cite{storms2022transparency, pennekamp2017survey}. Some participants argued that agency awareness should not belong exclusively to users but also to developers and regulators. They even framed this relationship as "adults" (platforms acting responsibly and simplifying the complexity) guiding "children" (users needing protection and support). Yet this analogy also exposes a core dilemma: even if platform-driven decisions lead to more moral or responsible outcomes, does treating users as "children" still strip away their agency? The challenge, then, lies in balancing establishment (risking paternalism) and autonomy (risking cognitive overload). Navigating this middle ground requires not only careful design but also a dedicated distribution of decision-making responsibilities across system actors.

\subsubsection{Transparency vs. User Comfort: The Paradox of Seeing More}
Our findings on interactive transparency surface an unresolved paradox: visibility into algorithmic processes is not always reassuring. Withholding information may simply not prompt users to think about underlying processes, leading to delight or frictionless experiences. Conversely, disclosing data or system mechanisms can build trust but also introduce anxiety and extra cognitive efforts to think, judge, and doubt, echoing previous studies \cite{kizilcec2016much, ananny2018seeing}. The visibility of options may also paradoxically heighten users’ awareness of what remains beyond their reach, thereby offsetting perceived gains in agency. This raises a challenge of what constitutes a good user experience: should systems reveal everything, or is selective disclosure preferable? Prior work suggests a balance of \textit{"not too little and not too much"} \cite{kizilcec2016much}, yet what constitutes sufficient visibility remains unclear. It may vary with users’ personalities, expectations, or tolerance for complexity, and may shift across platforms and contexts. As such, the ambivalence of "seeing more" reflects that transparency is not a straightforward variable to optimize but a paradox to navigate. As what paradox theory \cite{lewis2000exploring} describes as the simultaneous pursuit of contradictory logics: in this case, comfort through seamlessness and assurance through clarity, leaving design to negotiate trade-offs it may never fully resolve.

\subsubsection{Opacity vs. Agency? Navigating Platform Priorities}
Our study also highlights a broader strategic tension in RS development. Platforms often rely on opacity to create seamless, low-friction experiences to sustain engagement-driven business models. By hiding underlying processes, these systems reduce perceived complexity but also limit scrutiny and user influence. Many participants interpreted this lack of visibility as intentional, leading them to feel monitored or acted upon rather than engaged in a reciprocal relationship. Such feelings of surveillance and vulnerability reflect concerns about information and power asymmetries in algorithmic systems \cite{ruhr2023intelligent, kizilcec2016much, chen2018app}. Our study shows that transparency and meaningful controls can shift these dynamics. Even small points of insight or adjustability increased comfort, confidence, and, importantly, translate into willingness to engage with the platform \cite{chen2018app}. This suggests that agency-oriented designs may support business goals not by suppressing complexity, but by building trust-based engagement. This raises a strategic question for industry: should platforms continue to pursue engagement through inscrutability and unilateral control, or can empowering users cultivate more durable relationships? Our results indicate that agency does not compete with business priorities but can instead strengthen the platform–user relationship and function as an organizational asset. Reconciling these logics may require weighing short-term incentives against longer-term socio-technical considerations, inviting product strategies that treat user empowerment as a viable path to sustained engagement.

\section{LIMITATION AND FUTURE WORK}
This study has several limitations that point to opportunities for future research. First, the provotype was displayed only in a desktop format, and limited screen space (e.g., mobile) may constrain how features are presented or interpreted, requiring further design considerations. Second, although we used TikTok solely as a display environment and encouraged cross-platform thinking, the platform context may still have influenced how participants focused, recalled, or compared experiences, potentially limiting generalizability to other forms of AI RS. Relatedly, despite efforts toward diversity, the sample was primarily young adults with relatively high media and algorithm literacy, raising questions about applicability to groups with different ages (e.g., even younger high school students, undergraduates, or older adults), digital proficiency, or cultural backgrounds. Future work could incorporate a broader range to strengthen generalizability.

Furthermore, the provotype was non-executable, meaning participants formed impressions from short-term imagined interaction rather than from observing real system responses. While participants readily projected how features would alter their real feeds (e.g., mitigating filter bubbles), we did not empirically test whether the provotype would reduce those issues. Our focus was on whether such controls meet user needs and how participants understand and exercise agency when they are made available. As a result, observed shifts in perceptions and attitudes may not fully reflect real-world behavior. A natural next step is to examine how these features perform in real implementation and how they influence actual diversity and long-term personalization dynamics.

Another limitation is that our user controls were still pre-defined by the system. Although providing greater transparency and steerability than most existing interfaces, they cannot capture the full expressiveness of user intent as conversational RS do. Future work could compare fixed controls with more open-ended approaches to understand their respective impacts on user agency.

Beyond practical constraints, this study revealed valuable socio-technical tensions that merit further inquiry (see Section \ref{tension}). Future research can systematically investigate these tensions by identifying design strategies, gathering empirical evidence, and clarifying the unresolved trade-offs and considerations.

\section{CONCLUSION}
This paper investigated how interface-level interventions can enhance user agency in AI recommender systems. Motivated by concerns about information and power asymmetries, with filter bubbles as one example consequence, we designed a provotype that allowed users to actively manage data practices and navigate content discovery. Through provotype interactions and interviews, we investigated how these features influenced participants' experiences and attitudes toward these systems. Our findings show that visible algorithmic processes, clear action-outcome links, and accessible user controls can meaningfully shift recommender systems from opaque, prescriptive to more agency-oriented forms. Participants reported greater awareness of algorithmic processes, stronger perceived control, and improved trust when they could directly observe and influence recommendations. Overall, this work offers a generative foundation for refining transparency and control mechanisms in AI systems that balance usability, complexity, and responsiveness, and for reconciling algorithmic efficiency with user autonomy and meaningful participation.


\bibliographystyle{ACM-Reference-Format}
\bibliography{reference}

\appendix

\end{document}